\documentclass{article}
\usepackage{amssymb}
\usepackage{amsfonts}
\usepackage{amsmath}
\usepackage{graphicx}

\providecommand{\U}[1]{\protect\rule{.1in}{.1in}}
\DeclareMathOperator{\dive}{div}
\DeclareMathOperator{\rot}{rot}
\DeclareMathOperator{\rank}{rank}

\newtheorem{theorem}{Theorem}
\newtheorem{corollary}[theorem]{Corollary}

\newtheorem{example}[theorem]{Example}
\newtheorem{proposition}[theorem]{Proposition}
\newtheorem{remark}[theorem]{Remark}

\input{tcilatex}

\begin{document}

\title{On Poisson structures on $\mathbb{R}^{4}$}
\author{Rub\'en Flores-Espinoza \\
Departamento de Matem\'aticas,\\
Universidad de Sonora, M\'exico}
\date{Diciembre 2013}
\maketitle

\begin{abstract}
This paper is devoted to the study of Poisson structures on the Euclidean
four dimensional space $\mathbb{R}^{4}$. By using the properties of the
trace operator associated to a volumen form and the elementary vector
calculus operations in $\mathbb{R}^{3}$, we give explicit formulas for the
main geometric objects associated to the Poisson structures in $\mathbb{R}%
^{4}$, including its characteristic foliation, the Hamiltonian and Poisson
vector fields, normal forms and some useful decomposition formulae for
Poisson tensors. We also discuss the class of unimodular Poisson structures
and give two results about the existence of Poisson structures having as its
characteristic foliation a given arbitrary regular foliation.
\end{abstract}

\section{ Introduction}

In the last fifty years, Poisson geometry has been an important field of
research in differential geometry and mathematical physics for its relevance
in the mathematical foundations of classical and quantum mechanics. These
geometric structures are associated to the existence of contravariant
antisymmetric 2-tensors satisfying the so called Jacobi equation. The
Poisson tensors, or solutions of the Jacobi identiry, define a generalized
integrable vector field distribution in the sense of Sussman \cite{Su}
decomposing the total space into a family of symplectic manifolds fitting in
a smooth manner.

For low dimensional vector spaces, one can find in the literature several
papers devoted to the study of Jacobi equations, Poisson geometry and
Hamiltonian dynamics. For the tridimensional case, see for example, \cite%
{D-Z}, \cite{HB1}, \cite{HB}, \cite{LPV}, \cite{LX}. In dimension four, the
classification of linear and quadratic Poisson tensors has been done in \cite%
{K} and \cite{Sh}. In this article, we study from a general point of view
the Poisson structures on the four dimensional euclidean space $\mathbb{R}%
^{4}$. We give explicit formulas for the different quantities and structures
associated to a Poisson tensor on $\mathbb{R}^{4}$ by using the trace
operator properties and the elementary operations of vector calculus. In
particular, we study constant rank Poisson structures and those preserving a
given volume form. We discuss the automorphisms and infinitesimal symmetries
for Poisson tensors, focusing on the modular Poisson vector field and its
role on the preserving volume properties of Hamiltonian vector fields. We
include various useful decomposition formulas for Poisson tensors extending
some given results in the literature used to classify linear and quadratic
Poisson structures. Finally, we show that any regular distribution in the
sense of Frobenius corresponds to the symplectic foliation of some Poisson
structure. In particular, it the foliation is given by level sets of two
independent smooth functions, then the Poisson structure has a maximal basis
of infinitesimal automorphisms, transversal to the symplectic foliation.

The paper is organized as follows: In Section 2, we introduce the trace
operator on orientable manifolds, discussing its main properties and its
relation with the Schouten calculus for contravariant antisymmetric tensor
fields. We give explicit formulas for its action on the different
contravariant antisymmetric tensors in $\mathbb{R}^{4}$. In Section 3 we
introduce the fundamental concepts on Poisson structures and its particular
expressions on the vector space $\mathbb{R}^{4}$. We also derive some
explicit formulas for the different quantities associated to Poisson tensors
on $\mathbb{R}^{4}$, by using the elementary vector calculus operations for
three dimensional vector functions. Besides that, we present some important
classes of Poisson tensors. In Section 4, we study the group of symmetries
of a Poisson tensor and its infinitesimal generators, introducing the
modular vector field and the class of unimodular Poisson tensors. In Section
5 we give some interesting decomposition results for Poisson tensors in $%
\mathbb{R}^{4}$ by applying the particular properties of the trace operator
on orientable manifolds. Finally, in Section 6, we study the regular Poisson
structures on $\mathbb{R}^{4}$ and show that any regular 2-dimensional
distribution on $\mathbb{R}^{4}$, in the sense of Frobenius, is the
characteristic foliation of a regular Poisson structure. Furthermore, in the
particular case of foliations given by level sets of independent global
smooth functions, the Poisson structure possesses a maximal algebra of
transversal Poisson vector fields.\newline

\noindent\textbf{Acknowledgements}. This research was suggested by Professor
Yu. Vorobiev to whom I express my gratitude. I am grateful also to G.
Omelianov, J. Vallejo R., E. Velasco B., J.C. Ruiz P. and G. D\'{a}vila-Rasc%
\'{o}n for support and helpful discussions and comments on this work. This
paper was written under grant \ CB-178690 by Consejo Nacional de Ciencia y
Tecnolog\'{\i}a (M\'{e}xico).\newline

\section{The trace operator}

Let $M$ be a smooth, orientable $m-$manifold and $\Omega$ a volume form on $M
$. For $k=0,1,\cdots,m$ denote by $V^{k}(M)$ the space of antisymmetric
contravariant $k$-tensor fields on $M$. The \textit{trace operator} $\mathbf{%
D}$ \textit{relative to the volume form} $\Omega$, is an operator defined on
the space of contravariant tensor fields on $M$. For a $k-$tensor field $%
A\in V^{k}(M)$, $\mathbf{D}(A)$ is the unique $(k-1)-$tensor field defined
by 
\begin{equation}
\mathrm{d} \mathbf{i}_{A}\Omega = \mathbf{i}_{\mathbf{D}(A)}\Omega. 
\label{3.1.1}
\end{equation}
Here, $\mathbf{i}_{A}$ is the usual interior product operator on tensor
fields. The trace operator was introduced by Koszul in \cite{Kz} and it is
also called t he \emph{curl operator} or \emph{divergence operator} \cite%
{C-I-M-P}, \cite{D-Z}.

We have the following formulae for the different contravariant tensor fields
on $M$:

\begin{enumerate}
\item[(a)] If $X$ is a vector field on $M$, then $\mathbf{D}(X)$ is the
smooth function defined by the relation 
\begin{equation*}
\mathbf{D}(X)\Omega = \mathrm{d} \mathbf{i}_{X}\Omega = L_{X} \Omega. 
\end{equation*}

\item[(b)] For any vector fields $X$, $Y$ on $M$, we have 
\begin{equation}
\mathbf{D}(X\wedge Y)=[Y,X] +\mathbf{D}(Y)X-\mathbf{D}(X)Y   \label{3.1.4}
\end{equation}

\item[(c)] If $A \in V^{p}(M)$ and $B \in V^{q}(M)$, then 
\begin{equation}
\mathbf{D}(A\wedge B)=(-1)^{q}\mathbf{D}(A)\wedge B+A\wedge\mathbf{D}%
(B)-(-1)^{q+p}[A,B]   \label{3.1.5}
\end{equation}
where $[A,B]$ denotes the Schouten bracket between contravariant
antisymmetric tensors. Recall that for undecomposable $p$ and $q$-vector
fields on $M$, the Schouten bracket is defined by%
\begin{align*}
[ X_{1}\wedge \cdots\wedge X_{p},Y_{1}\wedge \cdots\wedge Y_{q}] & =
(-1)^{p+1}\sum_{i,j}(-1)^{i+j}[X_{i}, Y_{j}]\wedge X_{1}\wedge \cdots \\
& \cdots\wedge\widehat{X}_{i}\wedge\cdots\wedge X_{p}\wedge
Y_{1}\wedge\cdots\wedge\widehat{Y}_{j}\wedge\cdots\wedge Y_{q},
\end{align*}
where $\widehat{X}_{i}$ and $\widehat{Y}_{j}$ are deleted in each summand,
see \cite{Va}.

\item[(d)] For $A\in V^{p}(M)$ and $B\in V^{q}(M)$%
\begin{equation}
\mathbf{D}([A,B])=(-1)^{q} [\mathbf{D}(A),B] + [A,\mathbf{D}(B)]. 
\label{3.1.5.2}
\end{equation}

\item[(e)] The trace operator $\mathbf{D}$ is a cohomology operator%
\begin{equation}
\mathbf{D}^{2}=0.   \label{3.1.5.1}
\end{equation}
\end{enumerate}

If a $p$-tensor $A$ is in the kernel of the trace operator, we say that $A$
is a \textit{zero trace tensor}. In particular, the flow of a zero-trace
vector field preserves the volume form $\Omega$. In the case of a zero trace
2-tensor $\Lambda$, the vector field $\left[ f,\Lambda\right]$ is a
zero-trace vector field for each smooth function $f$. Moreover, the trace of
the Schouten bracket of zero-trace tensors is, again, a zero trace tensor.

If we consider another volume form $\widetilde{\Omega}=f \Omega$ with $f\neq0
$, then the corresponding trace operator $\mathbf{\widetilde{D}}$ is related
with $\mathbf{D}$ through the following relation 
\begin{equation}
\mathbf{\widetilde{D}(}A\mathbf{)=D(}A\mathbf{)-(-}1\mathbf{)}^{p}[\ln(\mid
f\mid),A], \qquad \forall \; A\in V^{p}(M).   \label{M.2}
\end{equation}

For more information about the trace operator see \cite{C-I-M-P}, \cite{D-Z}%
, \cite{K1}.

\begin{remark}
If $L= {\displaystyle \sum_{i\text{=1}}^{n}} x^{i}\frac{\partial}{\partial
x^{i}}$ is the $Euler$ vector field on $\mathbb{R}^{n}$ and $A$ is a $k$%
-homogeneous 2-tensor $A$ on $\mathbb{R}^{n}$, then, $[A,L]=(k-2) \, A$, and
the following decomposition formula holds 
\begin{equation}  \label{decompfor}
A =\frac{1}{n+k-2}(\mathbf{D}(A\wedge L)+\mathbf{D}(A)\wedge L).
\end{equation}
Decomposition formula (\ref{decompfor}) has been used in \cite{K} and \cite%
{Sh}, to classify quadratic and linear Poison tensors on $\mathbb{R}^{4}$,
respectively.
\end{remark}

Now, consider $\mathbb{R}^{4}$ with global coordinates $(\mathbf{x}%
,y)=(x_{1},x_{2},x_{3},y)$. In what follows, in order to write down the
different geometric quantities on $\mathbb{R}^{4}$, we will use the
operations $\times$, $\cdot$, $\nabla$, $\dive$, $\rot$ of elementary vector
calculus on the three variables $\mathbf{x}=(x_{1},x_{2},x_{3})$ and for any
smooth vector function $X = (X_1, X_2, X_3)$ on $\mathbb{R}^{3}$, we denote
by $X \frac{\partial}{\partial\mathbf{x}}$ the vector field 
\begin{equation*}
X \frac{\partial}{\partial\mathbf{x}} = X_1 \frac{\partial}{\partial x_{1}}
+ X_2 \frac{\partial}{\partial x_{2}} + X_3 \frac{\partial}{\partial x_{3}}, 
\end{equation*}
and by $X \frac{\partial}{\partial\mathbf{x}} \wedge \frac{\partial}{\partial%
\mathbf{x}}$ the $2$-tensor 
\begin{equation*}
X \frac{\partial}{\partial\mathbf{x}} \wedge \frac{\partial}{\partial\mathbf{%
x}} = X_1 \frac{\partial}{\partial x_{2}} \wedge \frac{\partial}{\partial
x_{3}} + X_2 \frac{\partial}{\partial x_{3}}\wedge\frac{\partial}{\partial
x_{1}} + X_3 \frac{\partial}{\partial x_{1}}\wedge\frac{\partial}{\partial
x_{2}}, 
\end{equation*}

With this notation we give the following formulae for the value of the trace
operator $\mathbf{D}$ on $\mathbb{R}^{4}$ relative to the canonical volume
form%
\begin{equation*}
\Omega= \mathrm{d}\mathbf{x}\wedge \mathrm{d} y = \mathrm{d} x_{1} \wedge 
\mathrm{d} x_{2} \wedge \mathrm{d} x_{3} \wedge \mathrm{d} y. 
\end{equation*}

\begin{enumerate}
\item[(a)] For a 4-tensor $A = f\frac{\partial}{\partial x_{1}}\wedge \frac{%
\partial}{\partial x_{2}}\wedge\frac{\partial}{\partial x_{3}}\wedge\frac{%
\partial}{\partial y}$, 
\begin{equation*}
\mathbf{D}A=\frac{\partial f}{\partial y}\frac{\partial}{\partial x_{1}}%
\wedge\frac{\partial}{\partial x_{2}}\wedge\frac{\partial}{\partial x_{3}}%
-\nabla f\frac{\partial}{\partial\mathbf{x}}\wedge\frac{\partial}{\partial%
\mathbf{x}}\wedge\frac{\partial}{\partial y}
\end{equation*}

\item[(b)] For a 3-tensor $A=g\frac{\partial}{\partial x_{1}}\wedge \frac{%
\partial}{\partial x_{2}}\wedge\frac{\partial}{\partial x_{3}}+\Sigma\frac{%
\partial}{\partial\mathbf{x}}\wedge\frac{\partial}{\partial \mathbf{x}}\wedge%
\frac{\partial}{\partial\mathbf{y}}$, we have%
\begin{equation*}
\mathbf{D}A=\left( \nabla g+\frac{\partial\Sigma}{\partial y}\right) \frac{%
\partial}{\partial\mathbf{x}}\wedge\frac{\partial}{\partial\mathbf{x}}-\rot%
(\Sigma)\frac{\partial}{\partial\mathbf{x}}\wedge\frac{\partial }{\partial y}
\end{equation*}

\item[(c)] For a 2-tensor $A=\Psi\frac{\partial}{\partial\mathbf{x}}\wedge 
\frac{\partial}{\partial\mathbf{x}}+\Phi\frac{\partial}{\partial\mathbf{x}}%
\wedge\frac{\partial}{\partial y}$, we have 
\begin{equation*}
\mathbf{D}(A)=\left( \rot(\Psi)+\frac{\partial\Phi}{\partial y}\right) \frac{%
\partial}{\partial\mathbf{x}}-\dive(\Phi)\frac{\partial}{\partial y}
\end{equation*}

\item[(d)] For a vector field $X=W\frac{\partial}{\partial\mathbf{x}} + b%
\frac{\partial}{\partial y}$%
\begin{equation*}
\mathbf{D}(X) = \dive(W)+\frac{\partial b}{\partial y}
\end{equation*}
\end{enumerate}

\section{Poisson structures on $\mathbb{R}^{4}$}

On $\mathbb{R}^{4}$ with global coordinates $(\mathbf{x}%
,y)=(x_{1},x_{2},x_{3},y)$, any antisymmetric contravariant 2-tensor $\Lambda
$ takes the form 
\begin{align}
\Lambda & =\Psi_{1}\frac{\partial}{\partial x_{2}}\wedge\frac{\partial }{%
\partial x_{3}}+\Psi_{2}\frac{\partial}{\partial x_{3}}\wedge\frac{\partial 
}{\partial x_{1}}+\Psi_{3}\frac{\partial}{\partial x_{1}}\wedge\frac{%
\partial }{\partial x_{2}}+  \label{1} \\
& +\Phi_{1}\frac{\partial}{\partial x_{1}}\wedge\frac{\partial}{\partial y}
+\Phi_{2}\frac{\partial}{\partial x_{2}} \wedge\frac{\partial}{\partial y}%
+\Phi_{3}\frac{\partial}{\partial x_{3}}\wedge\frac{\partial}{\partial y} , 
\notag
\end{align}
for smooth functions $\Psi_{i},\Phi_{i}$ on $\mathbb{R}^{4}$ and $i=1,2,3$.

If we consider the vector functions $\Psi=(\Psi_{1},\Psi_{2},\Psi_{3})$ and $%
\Phi=(\Phi_{1},\Phi_{2},\Phi_{3})$, we can identify the 2-tensor $\Lambda$
with the pair $(\Psi,\Phi)$ and write (\ref{1}) in the simplified notation%
\begin{equation}
\Lambda=\Psi\frac{\partial}{\partial\mathbf{x}}\wedge\frac{\partial}{\partial%
\mathbf{x}}+\Phi\frac{\partial}{\partial\mathbf{x}}\wedge \frac{\partial}{%
\partial y}.   \label{2}
\end{equation}

Each antisymmetric contravariant 2-tensor $\Lambda$ defines a morphism 
\begin{equation*}
\Lambda^{\#} : \bigwedge{}\negthickspace ^{1}(\mathbb{R}^{4})\rightarrow
V^{1}(\mathbb{R}^{4}) 
\end{equation*}
between the space of differential 1-forms and the space of vector fields on $%
\mathbb{R}^{4}$. The value of $\Lambda^{\#}$ on a 1-form $\alpha =A \, 
\mathrm{d} \mathbf{x} + b \, \mathrm{d} y$ on $\mathbb{R}^{4}$ is the vector
field 
\begin{equation}
\Lambda^{\#}(\alpha)=\left( \Psi\times A-b\Phi\right) \frac{\partial }{%
\partial\mathbf{x}}+A\cdot\Phi\frac{\partial}{\partial y}   \label{3}
\end{equation}
Any contravariant 2-tensor $\Lambda$ define a bilinear operation on the
space of smooth functions on $\mathbb{R}^{4}$ with values on the same space
of smooth functions given by the bracket%
\begin{equation}
\left\{ f,g\right\} = \mathrm{d} g(\Lambda^{\#}( \mathrm{d} f)).   \label{4}
\end{equation}
In terms of the vector functions $\Psi$ and $\Phi$ we have 
\begin{equation}
\left\{ f,g\right\} =\Psi\cdot\nabla f\times\nabla g + \Phi \cdot \left( 
\frac{\partial f}{\partial y} \nabla g - \frac{\partial g}{\partial y}
\nabla f\right).   \label{5}
\end{equation}

A contravariant antisymmetric 2-tensor $\Lambda$ is called a \textit{Poisson
tensor} or a \textit{Poisson structure} on $\mathbb{R}^{4}$ if the bracket (%
\ref{4}) satisfies the Jacobi identity 
\begin{equation}
\left\{ f,\left\{ g,h\right\} \right\} +\left\{ h,\left\{ f,g\right\}
\right\} +\left\{ g ,\left\{ h,f\right\} \right\} =0,   \label{6}
\end{equation}
for any smooth functions $f,g,h$ on $\mathbb{R}^{4}$. In this case, the
bracket (\ref{4}) is called the \textit{Poisson bracket} associated to the
Poisson tensor $\Lambda$. A 2-tensor $\Lambda$ satisfies the Jacobi identity
(\ref{6}), if and only if the vector functions $(\Psi,\Phi)$ are solutions
of the equation 
\begin{align}
& \Psi\cdot\left( \rot\Psi+\frac{\partial\Phi}{\partial y}\right) =\frac{%
\partial}{\partial y}(\Psi\cdot\Phi),  \label{7} \\
& \Phi\times\left( \rot\Psi+\frac{\partial\Phi}{\partial y}\right) +(\dive%
\Phi)\Psi =\nabla(\Psi\cdot\Phi).   \label{8}
\end{align}
Therefore, we can identify the Poisson tensors on $\mathbb{R}^{4}$ with
solutions of (\ref{7}), (\ref{8}).

In a general smooth manifold $M$, the Jacobi identity (\ref{6}), is
equivalent to the equation 
\begin{equation}
[ \Lambda,\Lambda] = 0.   \label{9}
\end{equation}
For orientable manifolds we have, from (\ref{3.1.5}), that a 2-tensor $%
\Lambda$ is a Poisson tensor if and only if 
\begin{equation}
\mathbf{D}(\Lambda\wedge\Lambda) = 2\Lambda\wedge\mathbf{D}(\Lambda). 
\label{3.1.2}
\end{equation}
Moreover, if $\Lambda$ is a 2-tensor on $M$ and we consider the $(m-2)-$form 
$\omega$ given by 
\begin{equation}
\omega = \mathbf{i}_{\Lambda}\Omega,   \label{10}
\end{equation}
then $\Lambda$ is a Poisson tensor if and only if 
\begin{equation}
\mathrm{d} (\mathbf{i}_{\Lambda}\omega) = 2 \mathbf{i}_{\Lambda} \mathrm{d}
\omega.   \label{11}
\end{equation}

Given a Poisson tensor $\Lambda$ on $\mathbb{R}^{4}$ and a smooth function $H
$, the vector field%
\begin{equation}
X_{H}=\Lambda^{\#}( \mathrm{d} H) = \left( \Psi\times\nabla H - \frac{%
\partial H}{\partial y} \Phi\right) \frac{\partial}{\partial\mathbf{x}} +
\nabla H \cdot \Phi \frac{\partial}{\partial y},   \label{H1}
\end{equation}
is called the \textit{Hamiltonian vector field} with \textit{Hamiltonian
function} $H$. Using the properties (\ref{3.1.5}) of the trace operator, the
Hamiltonian vector field $X_{H}$ satisfies%
\begin{equation*}
\mathbf{i}_{X_{H}} \Omega = - \mathrm{d} H \wedge \mathbf{i}_{\Lambda}
\Omega. 
\end{equation*}
In particular for Poisson structures on $\mathbb{R}^{4}$, the Hamiltonian
vector fields with Hamiltonian functions $H_{1} = - y$ and $H_{2} =
\Phi\cdot\Psi$ are, respectively, 
\begin{equation*}
X_{H_{1}} = \Phi\frac{\partial}{\partial x} \quad \text{and} \quad X_{H_{2}}
= (\Psi\cdot\Phi) \left( \left(\rot(\Psi) + \frac{\partial\Phi}{\partial y}
\right) \frac{\partial}{\partial\mathbf{x}} - \dive(\Phi)\frac{\partial}{%
\partial y} \right). 
\end{equation*}
In the case when $\dive(\Phi) = 0$, the Hamiltonian vector fields $X_{-y}$
and $X_{\Psi\cdot\Phi}$ commute.

The Jacobi identity for the Poisson bracket (\ref{4}) endows the space of
smooth functions with a Poisson algebra structure, which is homomorphic to
the Lie subalgebra of Hamiltonian vector fields 
\begin{equation}
[ X_{f},X_{g}] = X_{\left\{ f,g\right\} }, \qquad \forall \; f, g\in
C^{\infty}(\mathbb{R}^{4}).   \label{H3}
\end{equation}
Thus, the space of Hamiltonian vector fields define an integrable
generalized distribution of vector fields in the sense of Sussman \cite{Su}
and the total space $\mathbb{R}^{4}$ is foliated by leaves of (possible
different) even dimension. This foliation is called the \textit{%
characteristic foliation of} $\Lambda$ and each leaf $\mathfrak{L}$ is a
submanifold equipped with a symplectic structure given by the $2$-form 
\begin{equation}
\omega_{\mathfrak{L}}(X_{f},X_{g}) = \left\{ f,g\right\}\big|_{\mathfrak{L}%
}.   \label{H4}
\end{equation}
The characteristic foliation is also called the \textit{symplectic foliation}
of the Poisson structure.

Relative to a given Poisson tensor $\Lambda$ on $\mathbb{R}^{4}$, a smooth
function $k$ is called a \textit{Casimir function} if $\Lambda^{\#}(\mathrm{d%
} k) = 0$. Moreover, $k$ is a Casimir function if and only if 
\begin{equation}  \label{H5}
\left( \Psi\times\nabla k - \frac{\partial k}{\partial y} \Phi\right) =0
\quad\text{and} \quad \nabla k\cdot\Phi=0.
\end{equation}
The Poisson bracket of a Casimir function with any other smooth function
vanishes. Therefore any Casimir function is a first integral of any vector
field $\Lambda^{\#}(\omega)$, where $\omega$ is an arbitrary 1-form in $%
\mathbb{R}^{4}$.

In terms of the associated 2-form $\omega$ (\ref{10}) $k$ is a Casimir
function if and only if%
\begin{equation}
\mathrm{d} k \wedge\omega = 0.   \label{H6}
\end{equation}

The rank of the Poisson tensor $\Lambda = (\Psi,\Phi)$ is constant on the
points of each symplectic leaf. The classification of the symplectic leaves,
according to its dimension, is given by the following

\begin{proposition}
If $p = (\mathbf{x},y)\in\mathbb{R}^{4}$, the rank of a Poisson tensor $%
\Lambda=(\Psi,\Phi)$ at $p$, given by (\ref{2}), takes the following values 
\begin{equation*}
\rank\text{ }\Lambda(p)=\left\{ 
\begin{array}{lll}
0 & \text{ if and only if } & (\Phi^{2}+\Psi^{2})(p) = 0 \\ 
2 & \text{ if and only if } & (\Phi^{2}+\Psi^{2})(p)\neq0\text{ and }%
(\Phi\cdot\Psi)(p)=0 \\ 
4 & \text{ if and only if } & (\Phi\cdot\Psi)(p)\neq0%
\end{array}
\right. 
\end{equation*}
Moreover, the characteristic foliation of $\Lambda$ has the open sets 
\begin{equation*}
S^{+} = \left\{ (\mathbf{x},u)\in\mathbb{R}^{4} \; | \; \Phi\cdot \Psi > 0
\right\} 
\end{equation*}
and 
\begin{equation*}
S^{-} = \left\{ (\mathbf{x},u)\in\mathbb{R}^{4} \; | \; \Phi\cdot \Psi <
0\right\} , 
\end{equation*}
as its 4-dimensional symplectic leaves and its boundary 
\begin{equation*}
\partial S =\left\{ (\mathbf{x},u)\in\mathbb{R}^{4} \; | \; \Phi\cdot \Psi =
0 \right\}, 
\end{equation*}
is foliated by 2-dimensional and 0-dimensional symplectic leaves. Notice
that a point $p$ belongs to a singular symplectic leaf if $%
(\Phi\cdot\Psi)(p)=0$ and $\mathrm{d}(\Phi\cdot\Psi)(p) \neq 0$.
\end{proposition}

\begin{example}
The solutions of Jacobi equation with $\Phi=0$ define a parametrized family
of three dimensional Poisson structures where the parameter $y$ can be
considered as new variable commuting with all functions.
\end{example}

\begin{example}
The linear solutions $(\Psi,\Phi)$ of the Jacobi equation (\ref{7}), (\ref{8}%
) take the general form 
\begin{align}
\Psi(\mathbf{x},y) & = M\mathbf{x} + p\times\mathbf{x} + y\alpha,
\label{Li1} \\
\Phi(\mathbf{x},y) & = N\mathbf{x} + q\times\mathbf{x} + y\beta, 
\label{Li2}
\end{align}
where $M$ and $N$ are symmetric $3\times3$ matrices and $p,q,\alpha,\beta \in%
\mathbb{R}^{3}$ satisfy the equations 
\begin{align*}
& 2Mp=N\alpha-q\times\alpha \\
& 2\alpha\cdot p=\alpha\cdot\beta \\
& (2M+\Lambda\circ\beta)(N+\Lambda\circ q)+(N-\Lambda\circ q)(2M-\Lambda
\circ\beta)=2\,\mathrm{Tr}(N)M \\
& \mathrm{Tr}(N)\beta=N(2p+\beta)+q\times(2p+\beta) \\
& (\mathrm{Tr}N)\alpha-M\beta-p\times\beta = N\alpha - q\times\alpha.
\end{align*}
\end{example}

\begin{example}
The characteristic foliation of the linear Poisson bracket%
\begin{equation*}
\Lambda = 2x_{1}\frac{\partial}{\partial x_{2}}\wedge\frac{\partial}{%
\partial x_{3}} + \frac{1}{2}x_{1}\frac{\partial}{\partial x_{1}}\wedge\frac{%
\partial }{\partial y} + \frac{1}{4}x_{2}\frac{\partial}{\partial x_{2}}%
\wedge \frac{\partial}{\partial y} + \frac{1}{4}x_{3}\frac{\partial}{%
\partial x_{3} }\wedge\frac{\partial}{\partial y}
\end{equation*}
consists of:

\begin{enumerate}
\item[\upshape (a)] Two 4-dimensional symplectic leaves: $\mathcal{L}%
_{1}=\left\{ (\mathbf{x},y) \; | \; x_{1} > 0\right\} $ and $\mathcal{L}%
_{2}= \left\{ (\mathbf{x},y) \; | \; x_{1} < 0 \right\}$.

\item[\upshape (b)] The boundary $\left\{ (\mathbf{x},y) \; | \; x_{1}=0
\right\}$ is a submanifold foliated by 2-dimensional symplectic leaves of
the form $\left\{ x_{1}=0, \quad bx_{2}-cx_{3}=0,\quad bc \neq 0 \right\} $,
and zero dimensional leaves given by the points $(\mathbf{0},y)$.
\end{enumerate}

In particular, the 2-dimensional symplectic leaves are given by intersection
of the level set of $\displaystyle k = \frac{x_{2}}{x_{3}}$ with $x_{1}=0$.

The Hamiltonian vector fields take the form%
\begin{align*}
X_{f} & =\left( -2x_{1}\frac{\partial f}{\partial x_{3}}+\frac{1}{4}x_{2}%
\frac{\partial f}{\partial y}\right) \frac{\partial}{\partial x_{2}}+\left(
2x_{1}\frac{\partial f}{\partial x_{2}}\ +\frac{1}{4}x_{3}\frac{\partial f}{%
\partial y}\right) \frac{\partial}{\partial x_{3}} \\
& \qquad\qquad+\left( \frac{1}{2}x_{1}\frac{\partial f}{\partial x_{1}}+%
\frac{1}{4}x_{2}\frac{\partial f}{\partial x_{2}}+\frac{1}{4}x_{3}\frac{%
\partial f}{\partial x_{3}}\right) \frac{\partial}{\partial y}.
\end{align*}
\end{example}

If $\Lambda$ is a Poisson tensor on $\mathbb{R}^{4}$ has a smooth global
Casimir function $k : \mathbb{R}^{4} \rightarrow \mathbb{R}$, we have%
\begin{equation*}
\mathrm{d} k\wedge \mathbf{i}_{\Lambda}\Omega = - \mathbf{i}%
_{[k,\Lambda]}\Omega=0, 
\end{equation*}
and there exists a 1-form $\theta$ on $\mathbb{R}^{4}$ with%
\begin{equation}
\mathrm{d} k\wedge\theta = \mathbf{i}_{\Lambda}\Omega.   \label{C1}
\end{equation}
In this case, the Jacobi identity becomes%
\begin{equation}
\mathrm{d} \theta \wedge \theta \wedge \mathrm{d} k = 0.   \label{C2}
\end{equation}
Moreover, if we consider the 3-tensor $T$ given by 
\begin{equation*}
\theta = \mathbf{i}_{T}\Omega, 
\end{equation*}
then $\Lambda$ takes the form 
\begin{equation}
\Lambda = [k,T].   \label{C3}
\end{equation}
In global coordinates if $\theta = -A \mathrm{d} \mathbf{x} + f \mathrm{d} y$%
, with $A = A(\mathbf{x},y)$ a smooth vector function and $f$ a smooth real
function, then the Poisson 2-tensor $\Lambda$ in (\ref{C3}) takes the form 
\begin{equation}
\Lambda = \left( f\nabla k + \frac{\partial k}{\partial y}A\right) \frac{%
\partial}{\partial\mathbf{x}} \wedge \frac{\partial}{\partial\mathbf{x}} +
(A\times\nabla k) \frac{\partial}{\partial\mathbf{x}} \wedge \frac{\partial}{%
\partial y},   \label{K1.1}
\end{equation}
and the condition (\ref{C2}) becomes 
\begin{equation}
\left( \left( \nabla f + \frac{\partial A}{\partial y}\right) \times A - f
\; \rot(A)\right) \cdot \nabla k - \frac{\partial k}{\partial y} (A \cdot %
\rot(A)) = 0.   \label{k.1.1.1}
\end{equation}

Similarly, for those Poisson structures $\Lambda$ having 2 independent
global Casimir functions $k_{1}, k_{2}$ , we have%
\begin{equation*}
f \mathrm{d} k_{1} \wedge \mathrm{d} k_{2} = \mathbf{i}_{\Lambda}\Omega 
\end{equation*}
for some smooth function $f$. In this case, the Poisson tensor $\Lambda$ has
the form%
\begin{equation}
\Lambda = f\left( \frac{\partial k_{2}}{\partial y}\nabla k_{1}-\frac{%
\partial k_{1}}{\partial y}\nabla k_{2}\right) \frac{\partial}{\partial%
\mathbf{x}} \wedge \frac{\partial}{\partial\mathbf{x}} + f (\nabla
k_{1}\times \nabla k_{2}) \frac{\partial}{\partial\mathbf{x}}\wedge\frac{%
\partial}{\partial y}.   \label{K1.3}
\end{equation}

\begin{example}
The linear Poisson tensors $\Lambda$ having as Casimir function the
homogeneous polynomial of degree two $k_{1}=\frac{1}{2}\mathbf{x}^{T} P 
\mathbf{x} + (\alpha\cdot\mathbf{x}) y + \frac{1}{2} by^{2}$, with $P = P^{T}
$, $\alpha \in\mathbb{R}^{3}$ and $b \in\mathbb{R}$, take the form%
\begin{equation*}
\Lambda = ( c (P \mathbf{x} + y\alpha) + (\alpha \cdot \mathbf{x} + by) A) 
\frac{\partial }{\partial\mathbf{x}} \wedge \frac{\partial}{\partial\mathbf{x%
}} + (A\times ( P \mathbf{x} + y\alpha)) \frac{\partial}{\partial\mathbf{x}}
\wedge \frac{\partial }{\partial y}
\end{equation*}
where $A\in\mathbb{R}^{3}$, $c\in\mathbb{R}$. In this case, we have another
global Casimir function $k_{2}$ given by the linear function%
\begin{equation*}
k_{2} = A \cdot \mathbf{x} - c \; y 
\end{equation*}
\end{example}

For the general theory on Poisson structures, the reader can consult the
papers by A. Lichnerowicz \cite{LCH} and A. Weinstein \cite{We}, or the
books \cite{D-Z}, \cite{KaMs}, \cite{LPV}, \cite{Va}.

\section{Automorphisms of Poisson structures on $\mathbb{R}^{4}$}

Consider a Poisson tensor $\Lambda=$ $(\Psi,\Phi)$ on $\mathbb{R}^{4}$. A
diffeomorphism $F:\mathbb{R}^{4}\rightarrow\mathbb{R}^{4}$ with $F(\mathbf{x}%
,y) = (S(\mathbf{x},y),h(\mathbf{x},y))$, where $S:\mathbb{R}^{4}\rightarrow 
\mathbb{R}^{3}$ and $h:\mathbb{R}^{4}\rightarrow\mathbb{R}$ are smooth
functions, is called a \textit{Poisson map} if it preserves the Poisson
tensor or, equivalently, preserves the Poisson bracket%
\begin{equation}
\left\{ f, g\right\} \circ F = \left\{ f\circ F,g\circ F\right\}
,\qquad\forall \; f,g\in C^{\infty}(\mathbb{R}^{4}).   \label{1.6.0}
\end{equation}
In terms of the vector functions $(\Psi,\Phi)$ associated to the Poisson
tensor $\Lambda$, the condition (\ref{1.6.0}) is written as 
\begin{align}
\Psi(S(\mathbf{x},y),h(\mathbf{x},y)) & =\det(D_{\mathbf{x}}S)(D_{\mathbf{x}%
}^{-1}S)^{T} (\Psi) + D_{\mathbf{x}} S (\Phi) \times \frac{\partial S}{%
\partial y},  \label{1.6.1} \\
\Phi(S(\mathbf{x},y), h (\mathbf{x},y)) & = - D_{\mathbf{x}} S (\Psi\times
\nabla_{\mathbf{x}}h) + \frac{\partial h}{\partial y} D_{\mathbf{x}} S
(\Phi) - (\Phi\cdot\nabla_{\mathbf{x}}h)\frac{\partial S}{\partial y}. 
\label{1.6.2}
\end{align}
Here, $D_{\mathbf{x}}S$ denotes the differential of $S$ with respect to the
variable $\mathbf{x}=(x_{1},x_{2},x_{3})$.

Remark that expression (\ref{1.6.2}) is equivalent to 
\begin{equation}
\Phi(S(\mathbf{x},y), h (\mathbf{x},y)) = - DF(X_{h}),   \label{1.6.3}
\end{equation}
where $X_{h}$ is the Hamiltonian vector field corresponding to function $h(%
\mathbf{x},y)$ relative to the Poisson structure $(\Psi,\Phi)$. Notice that
the function $\Phi\cdot\Psi$ is transformed into 
\begin{equation*}
(\Phi\cdot\Psi) \circ F = (\det D F) (\Phi\cdot\Psi). 
\end{equation*}
In general, applying the transformation formulae (\ref{1.6.1}), (\ref{1.6.2}%
), the Poisson tensors having a global Casimir function can be transformed
to a normal form, as stated by the following result.

\begin{proposition}
Any Poisson tensor $\Lambda=(\Psi,\Phi)$ on $\mathbb{R}^{4}$ having a global
Casimir function $k$ with $\frac{\partial k}{\partial y} \neq0$ is
transformed, under the diffeomorphism $F : \mathbb{R}^{4} \rightarrow\mathbb{%
R}^{4}$, with $F(\mathbf{x},y) = (\mathbf{x}, k(\mathbf{x}, y))$, into the
normal form%
\begin{equation}
\widetilde{\Lambda} = \widetilde{\Psi} (\mathbf{x,}y) \frac{\partial}{%
\partial\mathbf{x}} \wedge\frac{\partial}{\partial\mathbf{x}}.   \label{K1.2}
\end{equation}
\end{proposition}

The constructions of Poisson brackets on smooth manifolds with prescribed
Casimir functions has been studied in \cite{D-P}.

Given a Poisson tensor $\Lambda$, a vector field $X=W\frac{\partial}{\partial%
\mathbf{x}}+b\frac{\partial}{\partial y}$ preserving the Poisson structure
is called an \textit{infinitesimal automorphism} of $\Lambda$ or \textit{%
Poisson vector field}. In terms of the Poisson bracket, $X$ is a Poisson
vector field if 
\begin{equation}
L_{X}\left\{ f,g\right\} =\left\{ L_{X}f,g\right\} + \left\{ f
,L_{X}g\right\} ,   \label{P0}
\end{equation}
for any smooth functions $f,g \in C^{\infty}(\mathbb{R})$, or equivalently, 
\begin{equation}
[ X,\Lambda] = 0.   \label{P00}
\end{equation}
In terms of the trace operator, condition (\ref{P00}) reads%
\begin{equation}
\mathbf{D}(\Lambda\wedge X)+\mathbf{D}(\Lambda) \wedge X -\mathbf{D}%
(X)\wedge\Lambda = 0.   \label{P01}
\end{equation}
Moreover, if $\Lambda$ is a Poisson tensor and $\omega=\mathbf{i}%
_{\Lambda}\Omega$, as in (\ref{10}), then a vector field $X$ is a Poisson
vector field if and only if 
\begin{equation}
L_{X}\omega = \mathbf{D}(X)\omega.   \label{P02}
\end{equation}
In $\mathbb{R}^{4}$ with global coordinates $(\mathbf{x,}y)$, the vector
field $X=W\frac {\partial}{\partial\mathbf{x}}+b\frac{\partial}{\partial y}$
is a Poisson vector field if and only if 
\begin{align}
\nabla(\Psi\cdot W)-(\dive\text{ }W)\Psi-W\times\rot\,\Psi+b\frac{\partial
\Psi}{\partial y}+\frac{\partial W}{\partial y}\times\Phi & =0,  \label{P1}
\\
\rot(\Phi\times W)-\dive(W)\Phi+\dive(\Phi)W+\Psi\times\nabla b-\frac{%
\partial b}{\partial y}\Phi+b\frac{\partial\Phi}{\partial y} & =0. 
\label{P2}
\end{align}
The space of Poisson vector fields is a Lie algebra of vector fields.

For each Poisson structure $\Lambda$ on an oriented manifold $(M,\Omega)$,
the vector field $Z_{\Lambda}=\mathbf{D}(\Lambda)$ satisfies 
\begin{equation}
[ Z_{\Lambda},\Lambda] = 0.   \label{3.1.2.0}
\end{equation}
Thus $Z_{\Lambda}$ is a Poisson vector field called the \textit{modular
vector field}. If we take on $M$ another volume form $\widetilde{\Omega}=f$ $%
\Omega_{0}$ with $f \neq 0$, the corresponding modular vector field $%
\widetilde{Z}_{\Lambda}$ differs from $Z_{\Lambda}$ by a Hamiltonian vector
field, 
\begin{equation*}
\widetilde{Z}_{\Lambda} - Z_{\Lambda} = X_{ - \ln(\mid f\mid)}. 
\end{equation*}

If $\Lambda$ is a Poisson tensor on $\mathbb{R}^{4}$ (\ref{1}), the modular
vector field $Z_{\Lambda}$ takes the form 
\begin{equation}
Z_{\Lambda}=\left( \rot(\Psi)+\frac{\partial\Phi}{\partial y}\right) \frac{%
\partial}{\partial\mathbf{x}}-\dive\Phi\frac{\partial}{\partial y}. 
\label{M.1}
\end{equation}

\begin{proposition}
\label{modtensfld} The modular tensor field $Z_{\Lambda}=\mathbf{D}(\Lambda)$%
, associated to a Poisson tensor $\Lambda$ on an orientable manifold $M$,
has the following properties:

\begin{enumerate}
\item[1.] For each Hamiltonian vector field $\ X_{H}$ we have: 
\begin{align}
\mathbf{D}(X_{H}) & =L_{Z_{\Lambda}}H,  \label{M1.1} \\
[ Z_{\Lambda},X_{H},] & =X_{\mathbf{D}(X_{H})}   \label{M1.2}
\end{align}
where $L_{Z_{\Lambda}}H$ denotes the Lie derivative of $H$ along $Z_{\Lambda}
$.

\item[2.] The Lie bracket of a Poisson vector field $W$ with the modular
vector field $Z_{\Lambda}$ is a Hamiltonian vector field with Hamiltonian
function given by the trace of $W$,%
\begin{equation}
[ W,Z_{\Lambda}]=X_{\mathbf{D}(W)}.   \label{M1.3}
\end{equation}
\end{enumerate}
\end{proposition}

From Proposition \ref{modtensfld} we deduce the following facts:

\begin{enumerate}
\item[(a)] If the modular vector field $Z_{\Lambda}$ vanishes, then any
Hamiltonian vector field has zero trace. Reciprocally, if all Hamiltonian
vector fields have zero-trace, then the modular vector field vanishes.

\item[(b)] From (\ref{M1.3}), the Lie bracket of $Z_{\Lambda}$ with any
other Poisson vector field $W$ is a Hamiltonian vector field. The trace of
any Poisson vector field commuting with the modular vector field $Z_{\Lambda}
$ is a Casimir function.
\end{enumerate}

For Poisson tensors on $\mathbb{R}^{4},$ the modular vector field $%
Z_{\Lambda}$ has the following additional properties:

\begin{enumerate}
\item[(a)] $Z_{\Lambda}$ is tangent to the level sets of $\Phi\cdot\Psi$.

\item[(b)] The formula $\Lambda^{\#}(\mathrm{d} (\Phi\cdot\Psi))=(\Phi\cdot%
\Psi)Z_{\Lambda}$ holds and for each Hamiltonian vector field with
Hamiltonian function $H,$ we have%
\begin{equation}
(\Phi\cdot\Psi)L_{Z_{\Lambda}}H=-L_{X_{H}}(\Phi\cdot\Psi).   \label{PVF2}
\end{equation}

Therefore, if $H$ is a first integral of the modular vector field $%
Z_{\Lambda}$, the Hamiltonian vector field $X_{H}$ is tangent to the level
set of the function $\Phi\cdot\Psi$. On the contrary, at any point $p\in%
\mathbb{R}^{4}$ where $(\Phi\cdot\Psi)(p)\neq0$ and $L_{Z_{\Lambda}}H(p)\neq0
$, the Hamiltonian vector field $X_{H\text{ }}$ is transversal to the level
set of $\Phi\cdot\Psi$.
\end{enumerate}

On a general orientable manifold $M$, a Poisson tensor $\Lambda$ is called 
\textit{unimodular} if for some volume form $\Omega$, the modular vector
field $Z_{\Lambda}$ vanishes. The unimodular Poisson tensors on $\mathbb{R}%
^{4}$ take the form%
\begin{equation}
\Lambda=\mathbf{D(}T)   \label{E.1}
\end{equation}
where $T$ is a 3-contravariant tensor 
\begin{equation}
T = f\frac{\partial}{\partial {x_1}} \wedge\frac{\partial}{\partial {x_2}}
\wedge \frac{\partial}{\partial {x_2}} + \Sigma\frac{\partial }{\partial%
\mathbf{x}}\wedge \frac{\partial}{\partial\mathbf{x}}\wedge \frac{\partial}{%
\partial {y}},   \label{E.2}
\end{equation}
where $\Sigma$ is a smooth vector function and $f$ a smooth function
satisfying 
\begin{equation}
\left( \nabla f +\frac{\partial\Sigma}{\partial y} \right) \cdot \rot%
(\Sigma) = c,   \label{E.3}
\end{equation}
with $c$ a constant.

Considering the 1-form $\theta$ defined by%
\begin{equation}
\theta = \mathbf{i}_{T}\Omega = - \Sigma \mathrm{d} \mathbf{x} + f \mathrm{d}
y,   \label{E.4}
\end{equation}
we have 
\begin{equation*}
\mathrm{d} \theta = \mathbf{i}_{\Lambda}\Omega, 
\end{equation*}
and condition (\ref{E.3}) is equivalent to 
\begin{equation*}
\mathrm{d} \theta \wedge \mathrm{d} \theta = c \, \Omega.
\end{equation*}
If the constant $c$ is non-zero, the unimodular Poisson tensor is
symplectic. If $c=0$, then $\Lambda$ is a rank 2, regular Poisson tensor.
The unimodular Poisson tensors $\Lambda$ on $\mathbb{R}^{4}$ take the form%
\begin{equation}
\Lambda=\left( \nabla f+\frac{\partial\Sigma}{\partial y}\right) \frac{%
\partial}{\partial\mathbf{x}}\wedge\frac{\partial}{\partial\mathbf{x}} - \rot%
(\Sigma)\frac{\partial}{\partial\mathbf{x}}\wedge\frac{\partial }{\partial y}%
.   \label{PL1}
\end{equation}

The unimodular Poisson tensors $\Lambda$ on $\mathbb{R}^{4}$ have the
following properties:

\begin{enumerate}
\item[(a)] The 2-form $\omega=\mathbf{i}_{\Lambda}\Omega$ is exact.

\item[(b)] The Poisson vector fields tangent to the symplectic foliation
have zero trace.

\item[(c)] The trace of a Poisson vector field transversal to symplectic
leaves is a Casimir function.

\item[(d)] For any 2-tensor $\Theta$ commuting with the unimodular Poisson
tensor $\Lambda$, the vector field $\mathbf{D(}\Theta\mathbf{)}$ is a
Poisson vector field.
\end{enumerate}

An example of unimodular Poisson tensors are those in $\mathbb{R}^{4}$ of
the form 
\begin{equation*}
\Lambda=\nabla G(\mathbf{x,}y)\frac{\partial}{\partial\mathbf{x}}\wedge 
\frac{\partial}{\partial\mathbf{x}}
\end{equation*}
Here, we have two global Casimir functions $k_{1}(\mathbf{x,}y)=y$ and $%
k_{2}=G(\mathbf{x,}y)$.

\begin{example}
In $\mathbb{R}^{4}$ with the canonical volume form, the quadratic Poisson
tensor \cite{K}%
\begin{equation*}
\Lambda = - x_{1}x_{2}\frac{\partial}{\partial x_{2}}\wedge\frac{\partial }{%
\partial x_{3}} + x_{1}x_{3}\frac{\partial}{\partial x_{3}}\wedge \frac{%
\partial}{\partial x_{1}} - yx_{1}\frac{\partial}{\partial x_{1}} \wedge%
\frac{\partial}{\partial y} - yx_{2}\frac{\partial}{\partial x_{2}} \wedge%
\frac{\partial}{\partial y} + yx_{3}\frac{\partial}{\partial x_{3}} \wedge%
\frac{\partial}{\partial y}
\end{equation*}
has a modular Poisson vector field given by%
\begin{equation}
Z_{\Lambda} = -2x_{1}\frac{\partial}{\partial x_{1}} - x_{2}\frac{\partial }{%
\partial x_{2}} + (2x_{3}+x_{1})\frac{\partial}{\partial x_{3}} + y\frac{%
\partial}{\partial y}.   \label{MVFE}
\end{equation}
In this case, we have two open 4-dimensional symplectic leaves given by $%
\mathcal{L} = \left\{ (\mathbf{x,}y) \; | \; yx_{1}x_{2} (x_{1} - x_{3}) > 0
(< 0) \right\} $. The set of rank-two points $(\mathbf{x,}y)$ is 
\begin{align*}
& \left\{ (\mathbf{x,}y) \; | \; yx_{1} = 0, \; y^{2} + x_{1}^{2}\neq 0,\;
x_{2}x_{3} \neq 0\right\} \cup \left\{ (\mathbf{x,}y) \; | \; yx_{1} \neq 0,
\; x_{2} = 0\right\} \\
& \qquad \cup \left\{ (\mathbf{x,}y) \; | \; yx_{1} \neq 0, \; x_{1} =
x_{3}\right\},
\end{align*}
and the zero dimensional leaves are given by the points $\left\{ (\mathbf{x,}%
y) \; | \; y^{2} + x_{1}^{2} = 0\right\} \cup \cup \left\{ (\mathbf{x,}y) \;
| \mathbf{x} = \mathbf{0} \right\} \cup  \left\{ (\mathbf{x,}y) \; | \;
x_{2} = x_{3} = y = 0\right\} $.
\end{example}

The unimodular Poisson structures has been studied by A. Weinstein in \cite%
{WE2}.

\section{Decomposition of Poisson structures}

In this section, we present some useful results on the decomposition of
Poisson brackets on $\mathbb{R}^{4}$as a sum of some special 2-tensors. We
recall that a pair of contravariant antisymmetric 2-tensors $A$ and $B$ is
called a \textit{Poisson pair} if both are commuting Poisson tensors, that
is $[A,A]=[B,B]=0$ and $[A,B]=0$. In this case, for any scalars $\lambda$, $%
\beta\in\mathbb{R}$, the 2-tensor $\lambda A+\beta B$ is again a Poisson
tensor.

\begin{proposition}
Let $\Lambda=(\Psi,\Phi)$ be a Poisson tensor on $\mathbb{R}^{4}$ and $%
\alpha = A \mathrm{d}\mathbf{x} + a \mathrm{d} y$ and $\beta = B \mathrm{d} 
\mathbf{x} + b \mathrm{d}y$ two independent 1-forms such that $\mathrm{d}%
\alpha\wedge\alpha \wedge\beta = \mathrm{d} \beta\wedge\alpha\wedge\beta=0$
and $\beta(\Lambda^{\#}(\alpha)) = 1$. Then $\Lambda$ has the following
decomposition%
\begin{equation}
\Lambda = \Lambda^{\#}(\alpha) \wedge \Lambda^{\#}(\beta) - (\Psi\cdot\Phi)
S_{(\alpha,\beta)},   \label{Fm1}
\end{equation}
where $S_{(\alpha,\beta)}$ is the rank-two Poisson tensor given by%
\begin{equation}  \label{Fm2}
S_{(\alpha,\beta)} = ( b A - a B) \frac{\partial}{\partial\mathbf{x}}\wedge 
\frac{\partial}{\partial\mathbf{x}} + (A\times B) \frac{\partial}{\partial 
\mathbf{x}} \wedge\frac{\partial}{\partial y}.
\end{equation}
In particular, if there exists independent smooth functions $f,g$ with $%
\left\{ f,g\right\} \neq 0$, then we have the decomposition formula 
\begin{equation}  \label{1.3.3.1}
\Lambda = \frac{1}{\left\{ f,g\right\} }X_{f}\wedge X_{g} - \frac{%
\Psi\cdot\Phi}{\left\{ f,g\right\} } S_{(\mathrm{d} f, \mathrm{d} g)},
\end{equation}
where%
\begin{equation}  \label{1.3.3.2}
S_{(\mathrm{d} f, \mathrm{d} g)} = \left( \frac{\partial g}{\partial y}%
\nabla f - \frac{\partial f}{\partial y}\nabla g\right) \frac{\partial}{%
\partial\mathbf{x}} \wedge \frac{\partial}{\partial\mathbf{x}} + \left(
\nabla f \times\nabla g \right) \frac{\partial}{\partial\mathbf{x}} \wedge 
\frac{\partial}{\partial y}.
\end{equation}
\end{proposition}

\begin{remark}
Notice that in (\ref{1.3.3.1}), the 2-tensor $S_{(\mathrm{d} f, \mathrm{d}g)}
$ is also a Poisson tensor and $S_{(\mathrm{d} f, \mathrm{d}g)}( \mathrm{d}%
f) = S_{( \mathrm{d}f, \mathrm{d}g)}(\mathrm{d} g) = 0$. Moreover, if a
Hamiltonian vector field $X_{f}$ is non-zero in some point $p$, then there
exists a function $h$ with $\left\{ f,h\right\} =1$, and $[X_{f},X_{h}]=0$
in some neigborhood of $p$. In this case, we have in the decomposition
formula (\ref{1.3.3.1}), that the tensor $X_{f}\wedge X_{h}$ is a Poisson
tensor. Thus, locally, we have a decomposition of $\Lambda$ as a sum of a
Poisson pair in the form 
\begin{equation}
\Lambda=X_{f}\wedge X_{h}-(\Psi\cdot\Phi)S_{(\mathrm{d} f, \mathrm{d} h)}. 
\label{1.3.3.3}
\end{equation}
and we recover the classical local decomposition theorem by Weinstein \cite%
{We}.
\end{remark}

Another type of decompositions of Poisson tensors on $\mathbb{R}^{4}$ are
the following:

\begin{proposition}
For each Poisson tensor $\Lambda=(\Psi,\Phi)$ on $\mathbb{R}^{4}$ , as in (%
\ref{1}), with $(\dive\Phi)\neq 0$, we have the decomposition%
\begin{equation}
\Lambda=Z_{\Lambda}\wedge\left( \frac{\Phi}{\dive(\Phi)}\right) \frac{%
\partial}{\partial\mathbf{x}}+\Lambda_{0},   \label{PS2}
\end{equation}
where 
\begin{equation*}
\Lambda_{0}=\frac{1}{\dive(\Phi)}\nabla(\Phi\cdot\Psi)\frac{\partial}{%
\partial\mathbf{x}}\wedge\frac{\partial}{\partial\mathbf{x}}
\end{equation*}
is a Poisson tensor of constant rank equal to 2.
\end{proposition}

\begin{corollary}
If the Poisson tensor is linear with $(\Psi,\Phi)$ given by (\ref{Li1}), (%
\ref{Li2}) and $\dive\Phi\neq 0$, then the decomposition formula (\ref{PS2})
can be rewritten as 
\begin{equation}
\Lambda = Z_{\Lambda}\wedge\Sigma\frac{\partial}{\partial\mathbf{x}}%
+\nabla(\Sigma\cdot\Psi)\frac{\partial}{\partial\mathbf{x}}\wedge \frac{%
\partial}{\partial\mathbf{x}},   \label{Lin3}
\end{equation}
with $\Sigma=\frac{1}{\dive\Phi}\Phi.$ In this case, the two 2-tensors $%
A=Z_{\Lambda}\wedge\Sigma\frac{\partial}{\partial\mathbf{x}}$ and $\
B=\nabla(\Sigma\cdot\Psi)\frac{\partial}{\partial\mathbf{x}}\wedge \frac{%
\partial}{\partial\mathbf{x}}$ are a Poisson pair.
\end{corollary}

For linear and quadratic Poisson tensors on $\mathbb{R}^{4}$, we recall the
following results given in \cite{Sh} and \cite{LLS}, \cite{LX}, respectively:

For linear Poisson tensor on a finite-dimensional vector space, applying
formula (\ref{3.1.5}) we have%
\begin{equation*}
\Lambda=\frac{1}{3}(\mathbf{D}(\Lambda\wedge L)+Z_{\Lambda}\wedge L) 
\end{equation*}
where $L$ denotes the \textit{Euler vector field} $L ={\displaystyle%
\sum_{i=1}^{n}} \mathbf{x}\frac{\partial}{\partial\mathbf{x}}+y\frac{\partial%
}{\partial y}.$

If the Poisson tensor $\Lambda$ is quadratic, then $[\Lambda,L]=0$ and
applying directly formula (\ref{3.1.5}) we obtain 
\begin{equation}
\Lambda=\frac{1}{4} \bigl( \mathbf{D}(\Lambda\wedge L) + Z_{\Lambda} \wedge
L \bigr).   \label{Q.1}
\end{equation}
In the quadratic case $L$ and $Z_{\Lambda}$ are commuting Poisson vector
fields of $\Lambda$. Moreover, $Z_{\Lambda}\wedge L$ and $\mathbf{D}(A
\wedge L)$ are commuting Poisson tensors and then, we have a decomposition
of $\Lambda$ as a sum of a Poisson pair, one of them with zero trace.
Applying the decomposition formula (\ref{Q.1}), the quadratic Poissson
tensors on $\mathbb{R}^{4}$ take the form%
\begin{equation*}
\Lambda = \mathbf{D}(T) + \frac{1}{4}X\wedge L, 
\end{equation*}
where $X$ is a linear vector field and $T$ is cubic 3-tensor on $\mathbb{R}%
^{4}$ satisfying the conditions%
\begin{equation}
\left[ \mathbf{D(}T\mathbf{),D(}T\mathbf{)}\right] =0,\qquad\mathbf{[D(}T),X%
\mathbf{]}=0, \qquad \mathbf{D}(X)=0.   \label{Q.3}
\end{equation}
If we consider the 1-form $\theta=\mathbf{i}_{T}\Omega$, conditions (\ref%
{Q.3}) can be written in the equivalent form%
\begin{equation}
\mathrm{d} \theta\wedge \mathrm{d} \theta = 0, \qquad \mathrm{d}(L_{X}
\theta) = 0, \qquad \mathbf{D(}X)=0,   \label{Q.4}
\end{equation}
Therefore, $\theta$ and $X$, satisfying (\ref{Q.4}), parametrize the
quadratic Poisson tensors on $\mathbb{R}^{4}$.

\section{Regular Poisson structures on $\mathbb{R}^{4}$}

A Poisson tensor is called \emph{regular} if it has a constant rank. It
means that its characterictic foliation is a regular foliation in the sense
of Frobenius. For regular Poisson tensors on $\mathbb{R}^{4}$ we have two
cases: one, when the rank of the Poisson tensor is 4 and the foliation
consists of only one leaf, and the other one, when each symplectic leaf is
two-dimensional.

The first case correspond to symplectic Poisson tensors. From Jacobi
conditions (\ref{7}) and (\ref{8}), a 2-contravariant tensor\textbf{\ }$%
\Lambda=(\Psi,\Phi)$ defines a symplectic structure on $\mathbb{R}^{4}$ if
and only if 
\begin{align}
(\Phi\cdot\Psi) & \neq0,  \label{S1} \\
\text{ }\dive \left( \frac{\Phi}{\Phi\cdot\Psi} \right) & =0,  \label{S2} \\
\frac{\partial}{\partial y} \left( \frac{\Phi}{\Phi\cdot\Psi} \right) + \rot %
\left( \frac{\Psi} {\Phi\cdot\Psi} \right) & = 0.   \label{S3}
\end{align}
Thus, from conditions (\ref{S1})-(\ref{S3}), the contravariant 2-tensor $%
\Lambda=(\Psi,\Phi)$ defines a symplectic structure on $\mathbb{R}^{4}$ if
and only if $\Phi\cdot\Psi\neq0$. In this case, the 2-tensor $\displaystyle 
\frac{1}{\Phi \cdot\Psi}\Lambda$ has zero trace and, on simply connected
domains, tensor $\Lambda$ takes the form $\Lambda=(\Phi\cdot\Psi)\mathbf{D}%
(T)$, where $T$ is some contravariant 3-tensor. Here, the modular vector
field $Z_{\Lambda}$ is the Hamiltonian vector field with Hamiltonian
function $H=\ln(\mid\Phi\cdot \Psi\mid)$.

Now, let us consider the Poisson structures $\Lambda$ on $\mathbb{R}^{4}$ of
rank equal 2. In this case, we have $\Lambda(\mathbf{x,}y)\neq0$, for all $(%
\mathbf{x,}y)\in\mathbb{R}^{4}$ and%
\begin{align*}
\Lambda\wedge\Lambda & = 0, \\
\Lambda\wedge\mathbf{D}(\Lambda) & = 0.
\end{align*}
In global coordinates $(\mathbf{x,}y),$ the regular Poisson tensors $\Lambda$
of rank 2 takes the form (\ref{1}) with $\Psi\cdot\Phi=0$ and $\Psi^{2}$ $%
+\Phi^{2}\neq0$. The Jacobi identity reduces to equations%
\begin{align*}
\Psi\cdot \left( \rot\Psi+\frac{\partial\Phi}{\partial y} \right) & = 0, \\
\Phi\times \left( \rot(\Psi)+\frac{\partial\Phi}{\partial y} \right) + (\dive%
\Phi)\Psi & = 0.
\end{align*}
Notice that if $\Lambda$ is a regular Poisson structure of rank 2, then for
any smooth function $\lambda\neq0,$ the tensor $\lambda\Lambda$ is also a
regular Poisson structure having the same characteristic foliation.
Moreover, for any smooth functions $f,g$ with $\left\{ f,g\right\} \neq0$ we
have the following representation, 
\begin{equation}
\Lambda=\frac{1}{\left\{ f,g\right\} }X_{f}\wedge X_{g}.   \label{I.8}
\end{equation}
In particular, in the open set $\Phi\cdot\mathbf{x}\neq0$ we have%
\begin{equation}
\Lambda = \left( \frac{-1}{\Phi\cdot\mathbf{x}} \right) X_{\left( \frac{1}{2}%
\mathbf{x}^{2}\right) }\wedge X_{y}.   \label{I.9}
\end{equation}

In the open domain where $\Phi\neq0,$ the regular Poisson tensor $\Lambda$
takes the form 
\begin{align}
\Lambda & =(\Phi\times\Sigma)\frac{\partial}{\partial\mathbf{x}}\wedge \frac{%
\partial}{\partial\mathbf{x}}+\Phi\frac{\partial}{\partial\mathbf{x}}\wedge%
\frac{\partial}{\partial y}  \notag \\
& =\Phi\frac{\partial}{\partial\mathbf{x}}\wedge \left( \Sigma\frac{\partial 
}{\partial\mathbf{x}} + \frac{\partial}{\partial y} \right),
\label{RegPoissTens}
\end{align}
for some vector function $\Sigma$ and the Jacobi identity reduces to the
single equation 
\begin{equation}
\Phi\frac{\partial}{\partial\mathbf{x}}\wedge \left[ \Sigma\frac{\partial }{%
\partial\mathbf{x}} + \frac{\partial}{\partial y},\Phi\frac{\partial} {%
\partial\mathbf{x}} \right] = 0.   \label{I.6}
\end{equation}

The Hamiltonian vector field $X_{H}$, at each point, belongs to the
distribution generated by the vector fields $\Phi\frac{\partial}{\partial%
\mathbf{x}}$ and $\Sigma\frac{\partial}{\partial\mathbf{x}}+\frac{\partial}{%
\partial y}$ and takes the form 
\begin{equation}
X_{H}=-\left( L_{\Sigma\frac{\partial}{\partial\mathbf{x}}+\frac{\partial }{%
\partial y}}H\right) \Phi\frac{\partial}{\partial\mathbf{x}}+\left( L_{\Phi%
\frac{\partial}{\partial\mathbf{x}}}H\right) \left( \ \Sigma \frac{\partial}{%
\partial\mathbf{x}}+\frac{\partial}{\partial y}\right) .   \label{H.2}
\end{equation}
The foliated symplectic form $\omega$ is%
\begin{equation}
\omega = \frac{1}{\Phi^{2} + (\Phi\times\Sigma)^{2}}(\Phi \mathrm{d} \mathbf{%
x)} \wedge (\Sigma \mathrm{d}\mathbf{x} + \mathrm{d} y),   \label{H.2.3}
\end{equation}
and the modular vector field $Z_{\Lambda}$ (\ref{M.1}) is given by%
\begin{equation}
Z_{\Lambda}=\left( \left[ \Sigma\frac{\partial}{\partial\mathbf{x}}+\frac{%
\partial}{\partial y},\Phi\frac{\partial}{\partial\mathbf{x}}\right] -\dive%
(\Phi)\left( \Sigma\frac{\partial}{\partial\mathbf{x}}+\frac{\partial }{%
\partial y}\right) +\dive(\Sigma)\Phi\frac{\partial}{\partial\mathbf{x}}%
\right).   \label{H.2.1}
\end{equation}

Taking into account (\ref{H.2.1}) we verify that $Z_{\Lambda}$ is always a
tangent Poisson vector field to the symplectic leaves. Notice that, on the
domain where $\dive(\Phi) \neq 0$, formula (\ref{H.2.1}) allow us to write
down the Poisson tensor (\ref{RegPoissTens}) in the form 
\begin{equation}
\Lambda=\frac{1}{\dive(\Phi)}Z_{\Lambda}\wedge\left( \Phi\frac{\partial }{%
\partial\mathbf{x}}\right).   \label{H.2.2}
\end{equation}

Any rank two regular foliation on $\mathbb{R}^{4}$ is generated by two
independent 1-forms $\alpha$, $\beta$, with $\alpha\wedge\beta \neq 0$, and
satisfying the integrability conditions 
\begin{align}
\mathrm{d} \alpha\wedge\alpha\wedge\beta & = 0,  \label{I.1} \\
\mathrm{d} \beta\wedge\alpha\wedge\beta & = 0.   \label{I.2}
\end{align}
In this case, the 2-tensor $\Lambda$ defined by the the relation 
\begin{equation}
\alpha\wedge\beta=\mathbf{i}_{\Lambda}\Omega,   \label{I.3}
\end{equation}
is a regular Poisson tensor.

Furthermore, using the local expressions for $\alpha,\beta$ given by $\alpha
= A \mathrm{d} \mathbf{x} + a \, \mathrm{d} y$ and $\beta = B \mathrm{d} 
\mathbf{x} + b \, \mathrm{d} y$, the 2-tensor field $\Lambda$ in (\ref{I.3})
takes the form 
\begin{equation}
\Lambda = (bA - aB) \frac{\partial}{\partial\mathbf{x}}\wedge\frac{\partial 
}{\partial\mathbf{x}} + (A\times B) \frac{\partial}{\partial\mathbf{x}}\wedge%
\frac{\partial}{\partial y},   \label{I.4}
\end{equation}
If $\alpha\wedge\beta$ is a closed $2$-form, then the modular vector $%
\mathbf{D}(\Lambda)$ vanishes and the Poisson structure $\Lambda$ is
unimodular.

We summarize the previous discussion in the following result.

\begin{proposition}
\label{prop:2rpt} Given a 2-dimensional integrable distribution $\mathfrak{F}
$ on $\mathbb{R}^{4}$ generated by two independent 1-forms $\alpha$ and $%
\beta$ satisfying (\ref{I.1})-(\ref{I.2}), there exists a Poisson structure $%
\Lambda$ having $\mathfrak{F}$ as its characteristic foliation.
\end{proposition}

For a regular foliation defined by the level sets of two independent
functions $f$ and $g$, we can improve Proposition~\ref{prop:2rpt}
constructing a Poisson structure which has the additional property of
possessing a basis of vector fields, transversal to the foliation. This kind
of srtuctures are called \textit{transversally constant Poisson structures} 
\cite{Va}.

Given two independent smooth functions $f$, $g$, we are going to construct a
Poisson structure with the above properties, in the following way: Take a
symplectic Poisson tensor $\Lambda = (\Psi,\Phi)$ on $\mathbb{R}^{4}$ with $%
\left\{ f,g\right\} \neq0$; then, consider the 2-tensor $\Delta$ given by 
\begin{equation}
\Delta =\Lambda-\frac{1}{\left\{ f,g\right\} }X_{f}\wedge X_{g}. 
\label{Dc1}
\end{equation}
The 2-tensor $\Delta$ in (\ref{Dc1}) is a regular Poisson having $f$ and $g$
as its Casimir functions and the vector fields%
\begin{equation}  \label{Dc2}
W_{1} =\frac{1}{\left\{ f,g\right\} }X_{f}, \qquad W_{2} =\frac{1}{\left\{
f,g\right\} }X_{g}, 
\end{equation}
as independent Poisson vector fields, transversal to its symplectic leaves.
The tensor $\Delta$ in (\ref{Dc1}) is called the \textit{Dirac--Poisson
tensor associated to $\Lambda$ and constraint functions $f$ and $g$}.

Taking into account the decomposition (\ref{1.3.3.1}), the Poisson tensor $%
\Delta$ takes the form 
\begin{equation*}
\Delta = -\frac{\Psi\cdot\Phi}{\left\{ f,g\right\} }S_{(\mathrm{d} f , 
\mathrm{d} g)}, 
\end{equation*}
and then, $\Delta$ is a regular Poisson structures having the desired
properties.

From the above discussion, given any regular foliation defined by the level
sets of two independent smooth functions $f$ and $g$ we can take any
symplectic structure $\Lambda$ and multiply the 2-tensor $S_{(\mathrm{d}f,%
\mathrm{d}g)}$ by the factor $\lambda = - \frac{\Psi\cdot\Phi}{\left\{
f,g\right\} }$ to have a transversally maximal Poisson structure whose
characteristic foliation coincides with the level sets of $f$ and $g$.
Notice that, in general, the constructed tensor $\Delta$ is not unimodular.

\begin{proposition}
For any regular foliation $\mathfrak{F}$ of rank 2, defined as the level
sets of two independent smooth functions $f$ and $g$, there exists a
transversally constant regular Poisson structure, having $\mathfrak{F}$ as
its characteristic foliation.
\end{proposition}


\begin{thebibliography}{99}
\bibitem{AGK} Ay A, Gurses M, and Zheltukin K., Hamiltonian equations in $%
\mathbb{R}^{3},$\textit{J. Math. Phys}. \textbf{44, } (2003) 5688-5705

\bibitem{C-I-M-P} Cari\~{n}ena J.F., Ibort, A., Marmo G., Perelomov A. On
the geometry of Lie algebras and Poisson tensors., \textit{J. Phys. A: Math.
Gen.} \textbf{27} (1994) 7425-7449.

\bibitem{D-P} Daminou A. P, and Petalidou F. Poisson brackets with
prescribed Casimirs.,arXiv:1103.0849v1 [Math. DG] 4 Mar 2011

\bibitem{DH} Dufour J.P. and Haraki A. Rotationnels et structures de Poisson
quadratiques, C.R. Acad. Sci. Paris Ser. I, Math. \textbf{312} (1991), no.1,
137-140

\bibitem{D-Z} Dufour J.-P. and Zung N.T. \textit{Poisson structures and
their normal forms, }Progress in Mathematics 242, Birkhauser Verlag, Basel
(2005)

\bibitem{GY} Gumral H., Nutku Y., Poisson structure of dynamical systems
with three degree of freedom. \textit{J. Math. Phys}. \textbf{34, }5691
(1993)

\bibitem{HB1} Hernandez-Bermejo B. New solutions of the Jacobi equations for
three dimensional Poisson structures, \textit{J. Math. Phys}. \textbf{42,}
(2001) 4984-4996

\bibitem{HB} Hernandez-Bermejo B., New solutions family of the Jacobi
equations: Characterization, invariants, and global Darboux analysis, 
\textit{J. Math. Phys.} \textbf{48 }(2007) 022903-022914

\bibitem{KaMs} Karasev, M. V. and Maslov, V. P., \textit{Nonlinear Poisson
brackets: Geometry and Quantization, }Translations of Mathematical
Monographs, V. 119, Amer. Math. Soc., Providence, 1993.

\bibitem{K} Klinker F., Quadratic Poisson structures in dimension four,%
\newline
www. Mathematik.uni-dortmund.de/\~{}klinker/Paper/PoissonQuadraticFour%
\_Klinker.pdf

\bibitem{K1} Klinker F., Polynomial Poly-vector fields, arXiv:math/0409157v4
[math. DG] Feb. 2008

\bibitem{Kz} Koszul, J.L., Crochets de Schouten-Nijenhuis et cohomologie, 
\textit{Asterisque, Soc. Math. de France, hors serie, (1985), 257-271}

\bibitem{LPV} Laurent-Gengoux C., Pichereau A. and Vanhaeecke P. \textit{%
Poisson Structures, }Springer Verlag Berlin Heidelberg 2013.

\bibitem{LCH} Lichnerowicz, A., Les varietes de Poisson et leur algebres
associes. \textit{J. Diff. Geometry. }\textbf{12 }(1977) 253-300

\bibitem{LLS} Lin, Q., Liu, Z.-J., and Sheng, Y.-H. Quadratic deformations
of Lie-Poisson structures, \textit{Lett. Math. Phys. }(2008) 83:217-299

\bibitem{Sh} Sheng. Y., Linear Poisson structures on $\mathbb{R}^{4}$. 
\textit{Journal of Geometry and Physics 57 (11) 2398-2410 (2007) }

\bibitem{Su} Sussmann, H.J., Orbits of families of vector fields and
integrability of distributions. Transactions Amer. Math. Soc. 170 (1973),
171-188

\bibitem{Va} Vaismann, I., \textit{Lectures on the Geometry of Poisson
Manifolds, }Birkhauser, Basel , 1994.

\bibitem{We} Weinstein, A., The local structure of Poisson manifolds, 
\textit{J. Diff. Geom. }\textbf{18 }(1983), 523-557

\bibitem{WE2} Weinstein, A., The modular automorphism group of a Poisson
manifold, J. Geom. Phys. \textbf{26 }(1997), 379-394

\bibitem{LX} Zhang Ju Liu and Xu Ping., On quadratic Poisson structures. 
\textit{Lett. Math. Phys.} \textbf{26 }(1) (1992), 33-42
\end{thebibliography}
\end{document}